\documentstyle[endfloat,epsfig,prl,aps]{revtex}

\begin{document}

\title{The Two Fluid Drop Snap-off Problem: Experiments and Theory}

\author{
Itai Cohen$^*$, Michael P. Brenner$^\dagger$,
Jens Eggers$^\ddagger$, Sidney R. Nagel$^*$
}

\address{
$^*$James Franck Institute, University of Chicago, Chicago, IL 60637 \\
$^\dagger$Department of Mathematics, MIT, Cambridge, MA 02139 \\
$^\ddagger$
Universit\"{a}t Gesamthochschule Essen, Fachbereich Physik,
45117 Essen, Germany \\ }

\maketitle
\begin{abstract}
We address the dynamics of a drop with viscosity $\lambda \eta$ breaking
up inside
another fluid of viscosity $\eta$. For $\lambda=1$, a scaling
theory
predicts the time evolution of the drop shape near the point of snap-off
which is in excellent agreement with experiment and previous simulations
of
Lister and Stone.  We also investigate the $\lambda$ dependence of the
shape and breaking rate.
\end{abstract}

When a fluid droplet breaks, as shown in Figure 1, a singularity develops
due to the infinite curvature at the point of snap-off \cite{E97}.  Near
such a singularity, the axial and
radial length scales become vanishingly small, allowing, independent of
initial conditions, a local analysis of the flow equations.  Such a
separation of scales implies that near snap-off the profiles should be
self-similar: on rescaling by the axial and radial scales the profiles
near
the singularity should collapse onto a universal curve.\cite{KM}

The character of the singularity depends on which terms in the
Navier-Stokes equations are dominant at the point of breakup.
If the drop breaks up in vacuum, surface tension,viscous stresses,
and inertia are balanced asymptotically, although the motion may
pass through other transient regimes, depending on viscosity
\cite{E93,Papa95,KM,BEJ,DHL}.  In
this paper, we investigate the situation where the viscous effects of the
inner and outer fluid are included as are the pressure gradients produced
by the
curvature in the surface separating them; the inertial terms are taken to
be insignificant so that we are in the Stokes regime \cite{LB98,LS98,L98}.
Assuming that molecular scales are not reached first, this is the final
asymptotic regime describing flows near snap-off for any pair of fluids
even in the case of arbitrarily low viscosity.  This paper uses
experiments, simulations and theory to characterize the self similar
approach to snap-off in this regime.
\begin{figure}
\centerline{\epsfig{figure=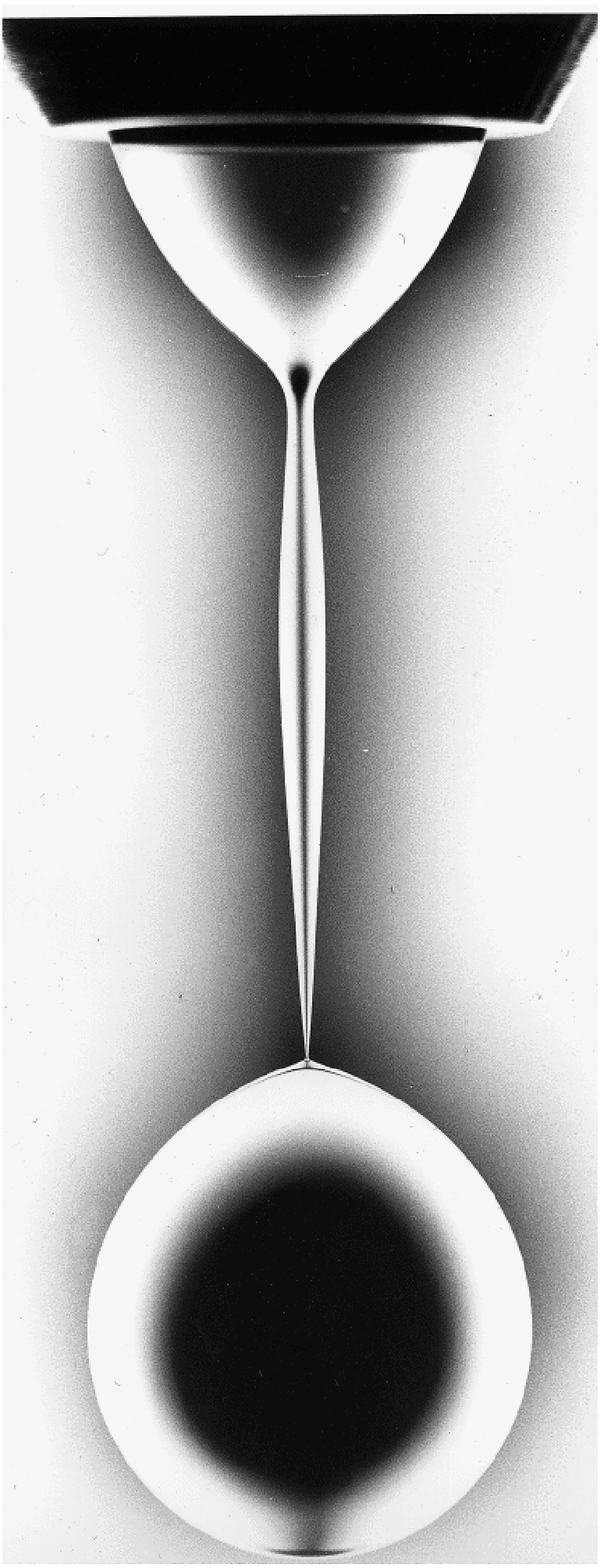}}
\caption[]{
A drop of 9.5 St Glycerin dripping through 10 St PolyDimethylSiloxane
(PDMS)
near snap-off. The nozzle diameter is $0.48$ cm.
}
\label{fig:drop}
\end{figure}

We consider the rupture of a fluid of viscosity $\lambda \eta$ surrounded
by
another fluid of viscosity $\eta$.  The interface between the two
fluids has surface tension $\gamma$.  At a time $t^*$ before the rupture,
dimensional analysis suggests
that all length scales have the form $H(\lambda)\gamma\eta^{-1} t^*$
where
$H(\lambda)$ is a function yet to be determined.  Hence, if drop profiles
near rupture are rescaled by $t^*$, they should collapse onto a universal
curve, independent of the initial conditions.  However, Lister and Stone
\cite{LS98} noticed that the long-ranged
character of the Stokes interaction leads to logarithmic corrections in
the
velocity field. They simulated equations
(\ref{velocity})-(\ref{interface})
below for drops having
various unstable initial conditions, and demonstrated collapse if the
logarithmic term was subtracted.  (See also Loewenberg et al. \cite{L98}.)

Herein, we demonstrate that this collapse also works for experiments, and
construct a scaling theory to explain the profile shapes, by incorporating
the nonlocal contributions into a local similarity description.  Figure 2
shows the collapse of rescaled profiles at $\lambda=1$ for both
experiments
and numerical simulations using a numerical technique similar to that of
Lister and Stone.  They are superimposed on the scaling theory developed
below (black line).
\begin{figure}
\centerline{\epsfig{figure=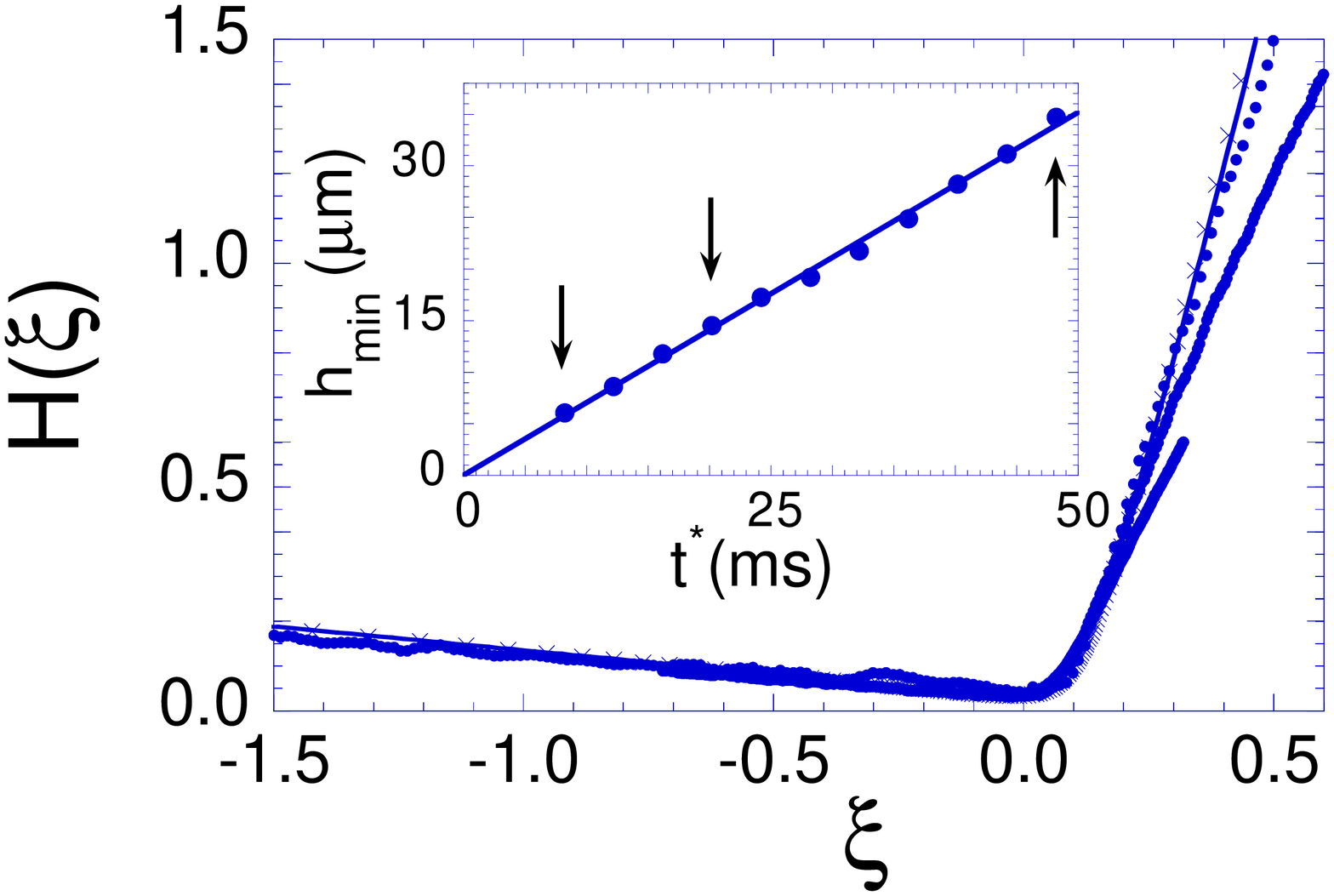,height=8cm}}
\caption[]{
The inset shows the minimum radius, $h_{min}(t)$, as a function of time
for
the drop shown in Fig. 1.  The solid line is the theoretical prediction.
The main figure shows the similarity function $H(\xi)$ as defined by
(\ref{similarity}).  The dots are rescaled experimental profiles
corresponding to the times  indicated as arrows in the inset. The solid
line is the theory, and the x's mark the final simulation profile.
}
\label{fig:similarity}
\end{figure}

The experiment used 9.5 St Glycerin dripping through 10 St PDMS.  The
viscosities are large enough that the experiment is in the Stokes-flow
regime even at macroscopic scales.
The surface tension $\gamma$ was measured using the pendant drop method
\cite{Neu} and the viscosity was measured using calibrated
Cannon-Ubbelohde
viscometers. We used a Kodak Motion Corder Analyzer to capture ten
thousand
frames per second. These images were then analyzed using an edge-tracing
program, and smoothed \cite{CN99}.

In rescaling the experimental profiles, we shifted the origin so that the
locations of the minima in the profiles lined up.  Because the profiles
were relatively flat along the axial direction there was some uncertainty
in the determination of these minima.  We therefore shifted each rescaled
profile in the axial, $\xi$, direction to minimize the cumulative
deviation
in $H(\xi)$.

The inset of Figure 2 shows that near snap-off $h_{min}(t)$ is a linear
function of $t^{*}$.  By fitting the prefactor of this linear dependence,
we obtain  $h_{min}=(0.031 \pm 0.008) \gamma \eta^{-1} t^*$, in excellent
agreement with the result $h_{min}= 0.0335\gamma \eta^{-1} t^*$ from
numerical simulations \cite{LS98} and the scaling theory constructed
below.

\underline{\bf Scaling Theory}
Since the Stokes equation is linear, the fluid surface velocity can be
expressed as
an integral over the surface of the fluid-fluid interface. At $\lambda=1$
the equation is\cite{RA78}
\begin{equation}
\label{velocity}
 {\bf v}^{(S)}(z,t) =
-\gamma \int \kappa(z') {\bf J}(z,z'){\bf n}(z') dz',
\end{equation}
where ${\bf n}$ is the outward normal, $\kappa$ is the curvature, $z$ is
the axial coordinate, and the tensor $\bf J$ is
\begin{equation}
\label{kernel}
{\bf J}(z,z') = \frac{1}{8\pi}\int_0^{2\pi}
\left[\frac{{\bf I}}{r} + \frac{{\bf r r}}{r^3}\right] d\theta
\end{equation}
with ${\bf r}$ the vector between the two points on the surface, $\bf{I}$
the identity matrix, and the integration is over the azimuthal angle
$\theta$ . Physically, equation (\ref{kernel}) represents the response of
the surface tension forcing the interface.  For unequal viscosities
$\lambda \ne 1$, eq. (\ref{kernel}) must be amended by an additional term,
which accounts for the jump in viscosity.  Given the radial $v_r$ and
axial
$v_z$ components of the surface velocity, the interface advances according
to
\begin{equation}
\label{interface}
\partial_t h(z,t) + v_z\partial_z h = v_r,
\end{equation}
which states that the surface at a given axial position can deform by
radial motion and axial advection.

Motivated by the simulations of Lister and Stone\cite{LS98}, we try the
similarity ansatz
\begin{equation}
\label{similarity}
h(z,t) = v_{\eta}t^{*} H(\xi) ,\; \xi = v_{\eta}^{-1} (z^{*}/t^{*})
+ b\ln t^{*} + \xi_0 ,
\end{equation}
where $z^*$ is the axial distance from the singularity, b is a constant,
and the factors of $v_{\eta}\equiv \gamma/\eta$ have been inserted to make
$H$ and $\xi$ dimensionless. The shift $b\ln t^{*}$ in the similarity
variable $\xi$ results from the logarithmic divergence of the axial
velocity field\cite{LS98}, and $\xi_0$ will be shown to be an arbitrary
constant which depends on the boundary conditions. Since the solution near
snap-off  must match onto the outer profile, which varies slowly on the
time-scale $t^{*}$
$H(\xi) \sim s_{\pm} \xi$, as $\xi\rightarrow\pm\infty$. Here we define
$s_-$ as the
slope of the shallow side of the pinch region, which by convention we
place to the 
left of the minimum, and $s_+$ as the steep slope.

The subtle feature of this problem is the interplay of the local
singularity with the nonlocal fluid response from the Stokes flow. The
principal nonlocal effect is that the surface tension force from the cones
produces a logarithmically diverging axial velocity field at the pinch
point, $\xi = 0$ \cite{LS98}. For a local scaling theory, this singularity
must be absorbed. We fix two points $\xi'_-$ and $\xi'_+$ within the
linear
part of the solution to the left and right of $\xi = 0$. Splitting the
contributions to the velocity on the surface into a contribution from
$\xi_- < \xi < \xi_+$ and from the rest of the drop, and converting to
similarity variables, we find
\begin{equation}
\label{vsim}
 {\bf V}^{(S)}(\xi,t^{*}) =
\left[-\int_{\xi'_-/t^{*}}^{\xi'_+/t^{*}} \kappa(\xi')
{\bf J}(\xi,\xi'){\bf n}(\xi') d\xi'- b\ln t^{*}{\bf e}_z\right]
,
\end{equation}
where ${\bf V}^{(S)} = ({\bf v}^{(S)}- b\ln t^{*}{\bf e}_z)/v_{\eta}$ and
${\bf e}_z$ is the unit
vector along the axial direction. Because of the cones, the axial
component of the J-integral in angular brackets diverges logarithmically
as $t^{*}\rightarrow 0$. For the special choice
$
b = -(s_+ (1+s_+^2)^{-1} + s_- (1+s_-^2)^{-1})/4$
the singularity cancels and the term in angular brackets remains finite
for $t^{*}\rightarrow 0$.
It is straightforward
to extend this scaling theory
to arbitrary $\lambda$; in
this case  the amplitude of $b$
only depends on  $\lambda$ through $s_+,s_-$.  The remaining constant in
(\ref{vsim}) depends on the detailed shape of the drop as well as on the
choice of $\xi'_-, \xi'_+$.

Inserting the similarity form (\ref{similarity}) into the equation of
motion for the interface (\ref{interface}) gives
\begin{equation}
\label{hequ}
-H + \frac{dH}{d\xi}(V_z + \xi - \xi_0 - b) = V_r,
\end{equation}
where we have absorbed the constant advection velocity $b \ln t^*$ into
$V_z$.

The system (\ref{vsim})-(\ref{hequ}) now has to be solved in the limit
$\xi'_{+} / t^{*} \rightarrow \infty$, $\xi'_{-} / t^{*} \rightarrow
-\infty$, for the interval $-\infty<\xi<\infty$ and with boundary
conditions $H\sim \xi$ as $\xi\to\infty$. Changing the constant $\xi_0
+ b$
only
results in a constant shift of the similarity function $H(\xi)$.  The
computation involves
solving an integro-differential equation with a nonlocal constraint: 
the parameter $b$ in (5) must be determined self consistently with the
solution $H(\xi)$ according to relation (\ref{vsim}). The difference
between this scaling theory and others developed for fluid rupture is that
here the {\sl parameters} in the similarity equation must be determined
self consistently with the solution to the similarity equation.

We solved this system by discretizing $H(\xi)$ in an
interval $\xi \in [-\xi_{in},\xi_{in}]$ and approximated all derivatives
and the integral by second-order formulas. At $\xi=-\xi_{in},\xi_{in}$
we demand $H''=0$. Using a linear approximation for $H$
outside the interval $[-\xi_{in},\xi_{in}]$, the logarithm
is subtracted explicitly. A numerical solution of the full PDE's
provided an initial condition for
Newton's iteration, which converged in a few steps. The iteration always
converges to the same solution for any given $\xi_0$.  The calculation
gives
$
H_0(\lambda=1) = 0.0335, \; s_- = -0.105, \; s_+ = 4.81,
$
where $s_{+,-}$ are the asymptotic slopes at $\pm\infty$. These results
are in good agreement with both the simulations of \cite{LS98}
and experiments (Fig. 2).  Although the
theory has been solved only for $\lambda=1$, by continuity we expect that
solutions exist for
a range of $\lambda$ and that $h_{min}$ obeys the law
$h_{min}= H_0(\lambda) v_{\eta} t^*.$

It is noteworthy that droplet breakup between fluids of equal
viscosities is not plagued by the iterated instabilities 
found for droplet breakup in vacuum \cite{instab}: neither
experiments nor simulations observe such 
instabilities. The reason for this can be found by
stability analysis of the similarity solution, following
the same procedure as \cite{BSN94}.  The result is
that perturbations around the similarity solution can
be amplified by a factor of $\approx 150$, which is much
smaller than the corresponding amplification factor for
rupture in vacuum, where it is $10^{4.7} (t^*)^{-1.5}$\cite{BSN94}.
The reasons for the differences between the two problems are that
(i) For $\lambda=1$ Tomitoka's formula \cite{T35} implies
that the  maximum linear
growth rate for perturbations is approximately $(t^*)^{-1}$.
On the other hand for $\lambda=\infty$,
the maximum growth rate is approximately $5 (t^*)^{-1}$.
(ii) The $\lambda=\infty$
problem has a time dependent amplification factor because the axial
length scale has a different scaling
with $t^*$ than the radial length scale.
This implies that the closest a perturbation can be to the
stagnation point depends on $t^*$, even when expressed in similarity
variables.

\underline{\bf Arbitrary $\lambda$}
We now extend features of the above results to arbitrary $\lambda$.  Using
Glycerin/water mixtures ($1$ St $< \eta <$ $9.5$ St) and silicone oils
($1$
St $< \eta <$ $600$ St) we were able to cover a range of $\lambda$ between
$0.002$ and $30$.
The same procedure as used above verified self-similar data collapse in
experiments with
$0.02 <\lambda < 30$\cite{CN99}.  In all these experiments, the conical
profile associated with $s_{-}$ collapsed for all analyzed profiles.
This corresponds to a time interval of at least $0.1$ seconds prior
to and all the way up to the point of snap-off. This can be seen in Figure
2
for $\lambda = 1$.  In contrast, as also seen in that figure, the conical
profile associated with $s_{+}$ showed a time-dependent collapse with the
region of self-similarity growing as the point of snap-off is approached.
This time dependence changes as we vary $\lambda$, slowing down as
$\lambda
\rightarrow 1$.

Figure 3 shows the cone angles $s_{\pm}$, and the dimensionless breaking
rate $H_0$ as a function of $\lambda$.  As shown by Figs. 3a-b, the cone
angles appear to obey a power law over an extended range of $\lambda$:
$s_+
\sim \lambda^{0.21\pm 0.05}$ and $s_- \sim \lambda^{-0.23\pm 0.07}$.  (Due
to measurement difficulties, the range for $s_-$ is reduced from that for
$s_+$\cite{CN99}.)  Within error, the analyses performed on both the
snap-off event near the nozzle and the snap-off event near the bulb lead
to
the same results. This agreement implies that the results are robust and
independent of small variations in the surrounding flows. Note that
our findings are in qualitative disagreement with lubrication-type
scaling arguments
\cite{LB98,LS98}, which predict $s_{\pm} \sim \lambda^{-1/2}$
for the slope on either side of the minimum.
On the other hand, the trends in our data are consistent
with recent numerical simulations of the full Stokes equations
by Zhang and Lister \cite{zha98}.  This is yet another indication 
of the breakdown of one dimensional models in describing
the dynamics of two fluid rupture.

\begin{figure}[h]
\centerline{\epsfig{figure=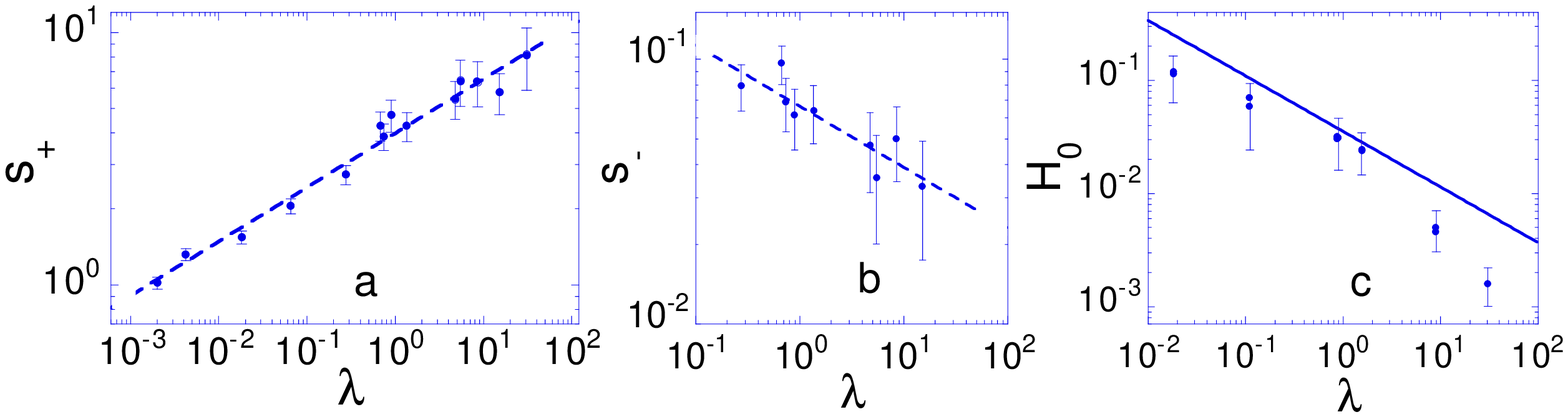,height=8cm,angle=0}}
\caption[]{\label{H}
The asymptotic slopes $s_+,s_-$, and the rescaled minimum radius
$H_0$ as a function of viscosity contrast $\lambda$. The dashed
lines are a fit to the experimental data,
in the right most graph the solid line is the result of our stability
argument.
      }
\end{figure}

\underline{\bf A Simple Theory for $H_0(\lambda)$}
follows by noting that
(assuming the shape of the drop
is slender near $h_{min}$)
the  maximum rate that the drop can break is given by the maximum linear
growth rate
$\Omega(h_{min})$ for a cylinder of uniform radius $h_{min}$.  Namely, we
have
the upper bound
\begin{equation}
\Omega(h_{min}) > \frac{\frac {d}{dt} (h_{min})}{h_{min}}=\frac{1}{t^*}.
\label{ub}
\end{equation}
By using Tomotika's formula \cite{T35} for $\Omega(h_{min})$ with
$h_{min}= H_0(\lambda) v_{\eta} t^*$, this equation turns into an upper
bound for $H_0(\lambda)$.
This upper bound is compared with the experimental data for $H_0(\lambda)$
in Fig. 3c. All of the  data obeys the bound; moreover, in the range
$0.1<\lambda<10$ the agreement is nearly exact. Note that while most of
the experimental data (and the upper bound equation (\ref{ub}) can be fit
with a power law of exponent $-0.53\pm  0.05$, a significant trend with an
overall negative curvature is observed in the experimental deviations.

The agreement of the experiments with the upper bound in the range
$0.1<\lambda<10$ is reminiscent of the marginal stability hypothesis, as
formulated for the
selection of traveling waves propagating from a stable to an unstable
state \cite{S89}.  Both experiments and numerical simulations show that
the breaking rate is approximately equal to the growth rate of linear
instabilities around a cylinder of radius $h_{min}$.
The upper bound in equation (\ref{ub}) should apply to all
problems involving singularity formation in a system with a local
instability.  We have tested this upper bound on similarity solutions from
several other examples including spherically symmetric gravitational
collapse\cite{lar69} and chemotactic collapse of bacteria\cite{bud}; the
upper bound is obeyed in each case,  giving a reasonable estimate for the
prefactor. Hence, this principle appears to be of rather general
applicability.

To conclude, we have (i) constructed a similarity solution for rupture at
$\lambda=1$, agreeing with previous numerical simulations \cite{LS98};
(ii) shown that experiments agree quantitatively with this similarity
solution, both in the form of the profile and its time dependence; and
(iii) presented a simple argument which quantitatively predicts the
breaking rate. Experiments have also shown self-similar behavior for the
range $0.02 <\lambda< 30$.
There are many unresolved issues: Among them, there
is no solid simple argument for the
$\lambda$-dependence of the slopes $s_-,s_+$.
Finally, our results
suggest that the scaling (\ref{similarity}) holds even in the limit
$\lambda \rightarrow \infty$, while a different set of scaling
exponents is found for a Stokes fluid breaking up in vacuum
$(\lambda = \infty)$ \cite{Papa95}. In addition, the profiles are
asymmetric for $\lambda \rightarrow \infty$, as also found in a recent
numerical simulation \cite{P98}, but are symmetric for breakup in
vacuum \cite{Papa95}, making this a singular limit.
Preliminary experimental results
of an inviscid fluid breaking up in a viscous fluid suggest that the
snap-off shape
is different from that in Stokes flow with $\lambda \rightarrow 0$
implying that this limit is singular as well.

We thank J. R. Lister, H. A. Stone, Q. Nie, L. P. Kadanoff, 
V. Putkaradze and T. Dupont for discussions. MB acknowledges
support from the NSF Division of Mathematical Sciences, and the A.P. Sloan
foundation. JE was supported by the Deutsche Forschungsgemeinschaft
through
SFB237. SRN and IC were supported by NSF DMR-9722646 and NSF MRSEC
DMR-9400379.

\end{document}